\documentclass[12pt]{article}        %%%%%%%%%%%%%%%%%%
\usepackage{epsfig}

\def\dek{$\Delta \kappa_\gamma$ }
\def\lam{$\lambda_\gamma$ }
\def\ib#1,#2,#3{       {\it ibid.\/ }{\bf #1} (19#2) #3}
\def\ap#1,#2,#3{       {\it Ann.~Phys.~(NY)\/ }{\bf #1} (19#2) #3}
\def\ijmp#1,#2,#3{     {\it Int.\ J.~Mod.\ Phys.\/ } {\bf A#1} (19#2) #3}
\def\mpl#1,#2,#3 {     {\it Mod.~Phys.~Lett.\/ } {\bf A#1} (19#2) #3}
\def\npb#1,#2,#3{       {\it Nucl.\ Phys.\/ }{\bf B#1} (19#2) #3}
\def\npps#1,#2,#3{     {\it Nucl.\ Phys.~B (Proc.~Suppl.)\/ }{\bf B#1}
                             (19#2) #3}
\def\plb#1,#2,#3{      {\it Phys.\ Lett.\/ }{\bf B#1} (19#2) #3}
\def\pr#1,#2,#3{       {\it Phys.\ Rev.\/ }{\bf #1} (19#2) #3}
\def\prd#1,#2,#3{      {\it Phys.\ Rev.\/ }{\bf D#1} (19#2) #3}
\def\prep#1,#2,#3{     {\it Phys.\ Rep.\/ }{\bf #1} (19#2) #3}
\def\prl#1,#2,#3{      {\it Phys.\ Rev.\ Lett.\/ }{\bf #1} (19#2) #3}
\def\pro#1,#2,#3{      {\it Prog.~Theor.\ Phys.\/ }{\bf #1} (19#2) #3}
\def\rmp#1,#2,#3{      {\it Rev.~Mod.~Phys.\/ }{\bf #1} (19#2) #3}
\def\sp#1,#2,#3{       {\it Sov.~Phys.~Usp.\/ }{\bf #1} (19#2) #3}
\def\zpc#1,#2,#3{      {\it Z.~Phys.\/ }{\bf C#1} (19#2) #3}
\def\appb#1,#2,#3{     {\it Acta Phys.\ Polon.\/ }{\bf B#1} (19#2) #3}
\def\cpc#1,#2,#3{      {\it Comput.\ Phys.\ Commun.\ }{\bf #1} (19#2) #3}
\begin{document}                     %%%%%%%%%%%%%%%%%%

%\allowdisplaybreaks

\begin{titlepage}

\begin{flushright}
CERN-TH/98-55\\
IFT/3/98
\end{flushright}

\vspace{1mm}
\begin{center}
{\Large\bf
Higher-order QED corrections to $e^+e^-\rightarrow \nu\bar{\nu}\gamma$ 
at LEP2
}
\end{center}
\vspace{3mm}

\begin{center}
{\bf A. Jacho\l kowska$^{1}$, J. Kalinowski$^{2}$ and Z. W\c as$^{3,4}$}\\

\vspace{5mm}
$^1${\em Laboratoire de l'Acc\'el\'erateur Lineaire,CNRS-IN2P3, 914050 ORSAY, France}\\
$^2${\em Institute of Theoretical Physics, Ho\.za 69, 00681 Warsaw, Poland}\\
$^3${\em CERN, Theory Division, CH-1211 Geneva 23, Switzerland}\\
$^4${\em Institute of Nuclear Physics,
  ul. Kawiory 26a, Cracow, Poland}
\end{center}

\vspace{10mm}
\begin{abstract}
  The process $e^+e^- \to \nu \bar\nu + \gamma$ with a distinctive
  ``photon-plus-missing-energy'' signal can serve as an important tool
  to search for new physics at LEP2.  It can be exploited to measure
  the $WW\gamma$ coupling, or to search for weakly interacting and
  invisible (s)particles.  For meaningful comparisons of experimental
  data with theoretical predictions, higher-order QED corrections due
  to multiphoton emission must be taken into account.  In the present
  paper we explain how the $WW\gamma$ coupling has been incorporated
  into the KORALZ Monte Carlo program, which can now be used for
  simulations of $e^+e^- \to \nu \bar\nu +\gamma$ events with
  higher-order QED corrections.  The strategy of how to perform an
  experimental analysis in the presence of experimental cuts is
  proposed. The question of systematic uncertainties is addressed and
  some numerical results for the phenomenologically interesting case
  of anomalous $WW\gamma$ couplings are also given.
\end{abstract}

\vfill
\begin{flushleft}
{ CERN-TH/98-55\\
     February 1998}
\end{flushleft}

\end{titlepage}

\centerline{\bf \large 1.  Introduction}
\vskip 0.3 cm

One of the main goals of experiments at high energy consists in
comparing new and so far unexplored measurable quantities with the
Standard Model (SM) predictions and thus providing its important tests
in a new experimental domain.  On the other hand, any discrepancy can
be a sign of new physics.  From the practical point of view it is thus
important to confront new experimental data with theoretical
predictions for observables which are: { (i)} sensitive to
expected new physics, { (ii)} under control of systematic
uncertainties for the data {\it and } theoretical predictions, and
{ (iii)} have high statistics.  To optimize the experimental
analysis and to suppress SM backgrounds, complicated patterns of
kinematical cuts are often introduced.  One should also bear in mind
that the detector sensitivity varies over detector acceptance in a
complicated manner.  That is why high-precision physics Monte Carlo
programs, including complete Standard Model predictions as well as
contributions from possible `new physics', are useful and in some
cases even mandatory for such projects \cite{cos}.

The measurement of the nature of three-gauge-boson couplings has been
identified as one of the most important topics in the LEP2 scientific
program \cite{ybook}. The most discussed $e^+e^-\rightarrow WW$
process in this context has a theoretical drawback as it involves two
$WWZ$ and $WW\gamma$ couplings intertwined, leading inevitably to a
model-dependent analysis.  In our study we will concentrate on the
phenomenology of a single $WW\gamma$ coupling. In \cite{ybook,wwg}
it was argued that the 
\begin{equation}
e^+e^- \to \nu \bar \nu \gamma
\label{nngamma}
\end{equation}
channel is a good candidate for such measurements, because the process
is sensitive only to this coupling, is of clean hadron-free signature
and has a rather large cross section.  Such final states have been
used at PETRA, SLAC and LEP1 \cite{nucount} as a means of measuring
the number of light neutrinos since, at these low energies, the
contribution from $WW\gamma$ coupling is negligible. However, at
higher energies the $t$-channel $W$-exchange process becomes
important, allowing for a direct study of $WW\gamma$ coupling,
independently of the $WWZ$ coupling. Therefore it opens an exciting
possibility since new physics can, and in general will, manifest
itself in corrections to the gauge-boson sector, in particular in the
$WW\gamma$ vertex.

The events with photon(s) plus missing energy in $e^+e^-$ collisions
might originate also from other mechanisms, signalling new physics
beyond the Standard Model. For example, such final states can be
produced in both gravity- and gauge-mediated supersymmetric models.
The missing energy in these events is caused by weakly interacting
supersymmetric particles, such as gravitinos, neutralinos and/or
sneutrinos \cite{others}.  In all such cases the Standard Model
$e^+e^- \to \nu \bar \nu \gamma$ events are irreducible background and
reliable predictions for them are therefore necessary.

Recently the process $e^+e^- \to \nu \bar \nu \gamma$ has been
exploited by the ALEPH collaboration 
to derive preliminary limits on anomalous $WW\gamma$ couplings from
data collected at 161, 172 and 183 GeV \cite{Aleph1}.  Although the results
for the couplings $\Delta\kappa_\gamma=0.05\  ^{+1.2}_{-1.1}$ (stat) and
$\lambda_\gamma=-0.05\  ^{+1.6}_{-1.5}$ (stat) are not competitive
with the ones expected to be derived (in a model-dependent way)  
from $WW$ final states at these energies, they
demonstrated the physics potential of processes with isolated photons.
Process (\ref{nngamma}) has also been used in searches for
supersymmetric signals \cite{Aleph2}.

To draw any conclusions on new physics from the experimental data,
however, a proper treatment of higher-order QED corrections to the
process in eq.~(\ref{nngamma}) is required.  In the present paper we will
discuss a Monte Carlo program that can be useful in reaching such a
goal.  As a specific example we consider the $\nu\bar{\nu}\gamma$
final states as a means to investigate the $WW\gamma$ coupling.  Our
paper is organized as follows. First we will present basic
distributions obtained from single bremsstrahlung analysis prior to
this work and recall that they were shown to be sensitive to anomalous
$WW\gamma$ couplings.  We will also show that they are prone to
higher-order QED corrections.  Then we will present the KORALZ Monte
Carlo program in which a realistic algorithm of how to match the
single-photon matrix element for $e^+e^- \to \nu \bar \nu \gamma$
(including anomalous $WW\gamma$ coupling) with higher-order QED
corrections has been implemented.  We will present consistency tests
of such an approach.  Finally we will turn to the question of the
choice of physical quantities that are most sensitive to anomalous
couplings when higher-order QED corrections are switched on.
Discussion of numerical results will conclude the paper.

\vskip 0.4 cm
\centerline{\bf \large 2.  $WW\gamma$ vertex and higher-order QED corrections}
\vskip 0.3 cm
Expressed in purely phenomenological terms, the effective Lagrangian
for the $WW\gamma$ interaction can be written in terms of seven
anomalous form factors \cite{hagi}. Assuming $C$ and $P$
conservation\footnote{Subject to possible cancellations between various
  contributions, data on the neutron electric dipole moment constrain
  these form factors very severely.}  reduces their number to three,
which are conventionally denoted by $\Delta g_1^{\gamma}$, $\Delta
\kappa_{\gamma}$ and $\lambda_{\gamma}$. The effective Lagrangian 
thus restricted can be written as follows
\begin{eqnarray}
  {\cal L}_{\it eff}^{WW\gamma}& = &
       -i e \bigg[ \hspace*{0.3em}
                         ( 1 + \Delta g^\gamma_1 )
                           ( W^\dagger_{\alpha \beta} W^\alpha
                                 - W^{\dagger\alpha} W_{\alpha \beta}
                           ) A^\beta
                         +
                          ( 1 + \Delta \kappa_\gamma)
                            W^\dagger_{\alpha} W_\beta A^{\alpha\beta}
       \nonumber         \\[1.5ex]
            &  & 
               + \frac{\lambda_\gamma}{M_W^2}
                 W^\dagger_{\alpha \beta} {W^\beta}_\sigma
                 A^{\sigma\alpha} \bigg]
\label{lagrangian}
\end{eqnarray}
where $A_{\alpha\beta} = \partial_\alpha A_\beta - \partial_\beta
A_\alpha $ and $W_{\alpha\beta} = \partial_\alpha W_\beta -
\partial_\beta W_\alpha $. All higher derivative terms can be absorbed
by the above couplings, provided they are treated as form factors and
not constants.

For the photons on-shell, as is the case in the process of interest 
$e^+e^- \to \nu \bar \nu \gamma$, electromagnetic gauge invariance
requires that $\Delta g^\gamma_1 (q^2 = 0) =0 $. Therefore we are left
with only two unknown form factors. In the Standard Model both couplings 
\dek and \lam vanish.

In the process $e^+e^- \to \nu \bar \nu \gamma$, there are only two
kinematic measurable variables at our disposal, for example the
energy and the polar angle of the observed photon. In order to achieve the
best sensitivity to anomalous couplings, it was shown in
Refs.~\cite{Aleph1,debchou} that double-differential distributions of
photons measured in the barrel detector are best suited. 
However, for the discussion of our
procedure it will be sufficient to refer to one-dimensional
differential distributions.

\begin{figure}[htb]
 \unitlength 1mm
\begin{picture}(100,100)
%\lattice
  \put(20,-10){ \epsfxsize=10cm \epsfysize=15cm
\epsfbox{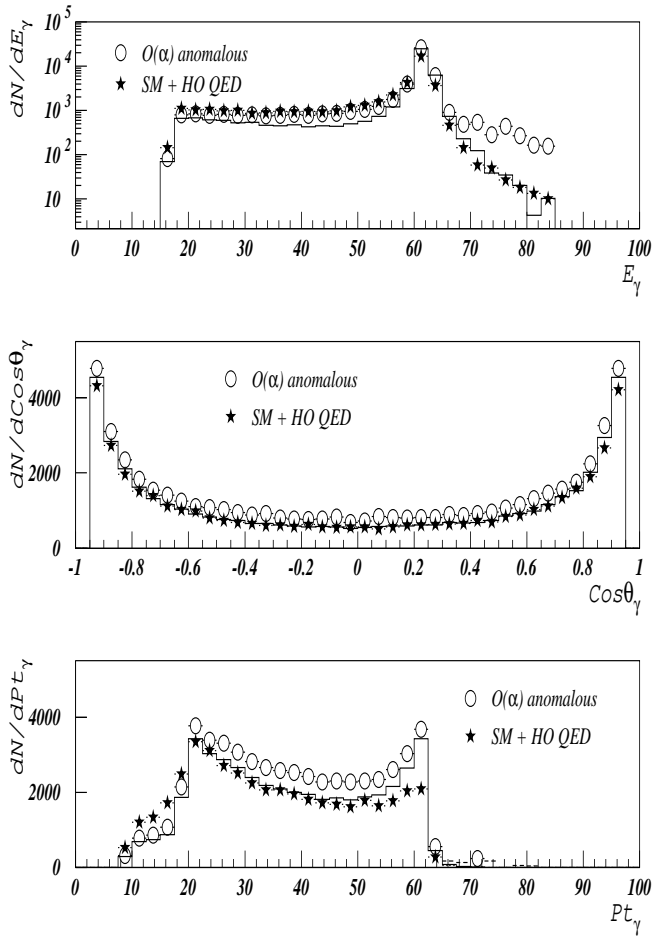}}
\end{picture}
%\begin{center}
%  \epsfig{file=geng3.eps,width=10cm,height=12cm} 
\caption{Inclusive photon distributions for: 
   (i)~genuine order-$\alpha$ distributions in the Standard Model
  -- solid line; (ii)~order-$\alpha$ including anomalous interactions
  -- open circles; and (iii)~the SM with multiphoton radiation -- no
  $WW\gamma$ interaction -- stars. For the set of experimental cuts and
  anomalous couplings see the text.}
\label{fig:classical}
%\end{center}
\end{figure}

We use the following set of selection criteria, which have been used
by ALEPH: (i)~photon transverse momentum $p_T> 0.05 \sqrt{s}$,
(ii)~photon energy $E_\gamma > 0.1 \sqrt{s}$, (iii)~photon angular
acceptance region $|\cos \theta_\gamma | < 0.95$, and for definiteness
we take the centre-of-mass energy $\sqrt{s}=172$ GeV. In
Fig.~\ref{fig:classical} the energy, the angular and the transverse
momentum distributions for a single photon are shown for three cases:
genuine ${\cal O}(\alpha)$ SM results (solid lines), SM including
multiphoton radiation but with $WW\gamma$ interaction exluded (stars),
and genuine ${\cal O}(\alpha)$ including anomalous $WW\gamma$ coupling
(open circles). The maximum around $E_r=60$~GeV in the energy
distribution corresponds to the so-called radiative return to the $Z$,
{\it i.e.} the photon energy for which the invariant mass of the
recoiling system $\nu \bar \nu$ pair is close to the $Z$ mass.

There are several comments to be made. Comparing solid lines with
stars, it is seen that photons with energies higher than $E_r$ are best
suited for the $WW\gamma$ coupling measurement, because of a sizeable
sensitivity to anomalous contributions.  For the energies matching
radiative return, the significance is very small because of the high
Standard Model background. For energies below $E_r$ the photons are
also sensitive to anomalous couplings, although the sensitivity
decreases systematically with decreasing photon energy.  Below 20 GeV
the sensitivity is negligible even to rather strong couplings, which
we have taken in the calculations, namely $\Delta \kappa_{\gamma} =
 - 10$, $\lambda_\gamma=0$. We will use this set of anomalous couplings
throughout the paper as an example.
The above conclusions are derived at the single-photon emission level.
This might be sufficient to estimate the discovery power of the
appropriate observables. However, it must be completed with a careful
analysis of higher-order QED corrections because these may change not
only the size but also the shape of the distributions that are
analysed.  To illustrate this point, in Fig.~\ref{fig:classical} we
include for comparison the results of a multiphoton higher-order Monte
Carlo simulation performed with the help of the KORALZ program
\cite{Koralz}, in which the $WW\gamma$ coupling is not included (open
circles).  For photons of energies higher than that of radiative
return, the effect of higher-order QED corrections is small; they
slightly decrease the SM cross section.  In that kinematical region
the inclusion of higher-order correction does not seem to be essential
for our choice of the anomalous couplings.  The situation is different
for photon energies below that for radiative return. Contributions
from both anomalous couplings and higher-order QED corrections to the
distributions are of similar order and have the same sign (for the
values of \dek and \lam chosen).  Higher-order corrections vary in
that region from $-30$\% to about 90\%. This result shows that any
realistic determination of the anomalous $WW\gamma$ coupling must
compare data with theoretical predictions that include anomalous terms
as well as higher-order QED corrections at the same time. This
observation is the main motivation for our present work.
 
\vskip 0.4 cm
\centerline{\bf\large 3. Monte Carlo generator}
\vskip 0.3 cm

The most up-to-date published description of the KORALZ 4.02 program for
production of pairs of leptons, including radiative QED corrections, is
described in detail in Ref.~\cite{Koralz}.  Real photon radiation
can be generated at different levels of sophistication.  First, the
program may be run at the Born level for the $e^+e^- \to l \bar l$
process\footnote{ This option is not interesting for $l=\nu$ since no
  photon radiation is present.}.  Second, it may be run at order
$\alpha$ when single bremsstrahlung configurations are generated.
Finally, the program can be run with pragmatic order-$\alpha^2$ QED
corrections including exclusive exponentiation.  In this case the
number of photons actually generated is not restricted.

\begin{figure}[htbp]
        \vskip 4in\relax\noindent\hskip -1.8in
        \relax{\includegraphics{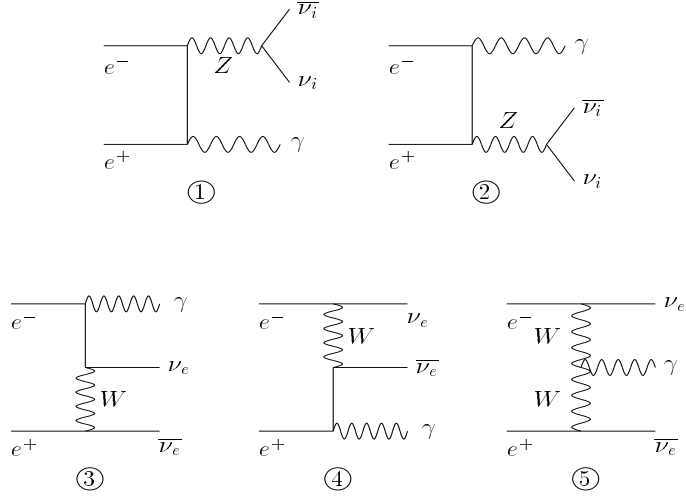}}
        \vspace{-20ex}
 \caption{{\em The Feynman diagrams for
          $e^+ e^- \rightarrow \bar \nu \nu \gamma$.}}
 \label{feyn}
\end{figure}

Let us briefly recall the main features of KORALZ existing 
prior to the present work, which are relevant for
$e^+e^- \to \nu \bar \nu \gamma $ production.  The basic design of
KORALZ is for the $s$-channel processes, such as muon or tau lepton
pair production.  The cases of muon and tau neutrinos accompanied by a
single photon, see diagrams (1) and (2) in Fig.~\ref{feyn}, have been
implemented easily, since the modification is straightforward,
consisting of changing the appropriate coupling constants only. In the
case of electron neutrinos, there are additional contributions from
$t$-channel $W$ exchange, diagrams (3) and (4), and from the
$WW\gamma$ vertex, diagram (5) in Fig.~\ref{feyn}.  Since in physical
gauge the amplitude for the diagram (5) with $WW\gamma$ vertex is
finite in soft and collinear limits and negligible at LEP1
\cite{berends}, it was not implemented at the time.  Note that by
neglecting the $WW\gamma$ vertex the generated distributions are
formally not gauge-invariant.
%
%%%%%%%%                   Feynman Diagrams for nu nu gamma
%%%%%%%%%%%%%%%%%%%%%%%%%%%%%%%%%%%%%%%%%%%%%%%%
%
Diagrams (3) and (4) have been implemented in an approximate
way\footnote{ It is thus important to realize that the program quality
  for $\nu_e \bar \nu_e \gamma $ final state was (and remains)
  inferior to that for other final states.} as follows.  First a series
of (weighted) events is generated inside KORALZ according to the 
Born-level $s$-channel $Z$-exchange diagrams for neutrino pair production
(leading to totally unobservable final states), and with higher-order
QED corrections including additional virtual and/or real photons.  In
this way final states consisting of 0, 1, 2, 3, ..., photons
accompanying the neutrino pair are generated according to the
prescription of exclusive exponentiation of YFS \cite{YFS}.  Full
phase space is covered by the generation. At this point the complete
event with four momenta of final-state neutrinos and photons is
constructed and stored in the common block. This procedure is correct
for $\nu_\mu \bar \nu_\mu n\gamma $ final states. For the electron
neutrinos, the matrix element for the hard process $e^+e^-\rightarrow
\nu_\mu \bar \nu_\mu$ (with $Z$ exchange only) has to be replaced by
the matrix element for $e^+e^-\rightarrow \nu_e \bar \nu_e$ (with $Z$
{\it and} $W$ exchanges).  Consider first ${\cal O}(\alpha)$
contributions coming from $Z$ (1+2) and $W$ (3+4) diagrams.  Thanks to
the factorization properties of the QED, the
matrix element squared $|{\cal M}_{0}|^2$ for the sum of four
diagrams (1+2+3+4) is approximated as a product of the matrix element
squared $|{\cal M}_{12}|^2$ for two diagrams (1+2) and the weight
factor $F_W$:
\begin{eqnarray}
|{\cal M}_{0}|^2 &\simeq&
| {\cal M}_{0}^{approx}|^2 =|{\cal M}_{12}|^2 F_W,
       \nonumber         \\[1.5ex]
F_W&=&{| {\cal M}_{Z}+ {\cal M}_{W}|^2 \over |{\cal M}_{Z}|^2}, 
\label{Factor}
\end{eqnarray}
where ${\cal M}_{Z}$ and ${\cal M}_{W}$ denote amplitudes for $\nu_e
\bar \nu_e $ pair production via $Z$- and $W$-exchange diagrams,
respectively, with {\it no photons} in the final state.  In the
calculation of the factor $F_W$ four momenta of $\nu_e $ and $\bar
\nu_e $ are used in a straightforward way, but for the incoming
electrons and positrons effective beam momenta are calculated from
momenta of the real beams and real photon emitted, according to the
prescription given in Refs.~\cite{Koralz,colas1}.  Note that the
factor $F_W$ given in eq.~(\ref{Factor}) is gauge-invariant. 
The standard final rejection of KORALZ creates the sample of
unweighted events, which is passed to the program user. The factor
$F_W$ is included in that final rejection weight.

Let us stress that in this way the effect of diagrams (3) and (4) with
$t$-channel $W$ exchange, as well as their interference with the
diagrams (1) and (2), has been introduced not
only for case of a single-photon radiation, but also for multiphoton
configurations. This approximation, as well as the whole
program, has been heavily tested at LEP1 energies  both for the
single bremsstrahlung mode and the multiple photon case.  Extensive
comparisons with the results from another program \cite{NNGG}, including
a complete matrix element for single and double bremsstrahlung, have
been performed, finding an excellent agreement and thus proving both
the validity of the approximation and the correctness of the code,  see
Refs.~\cite{Koralz,colas1,colas} for more details.
During the last seven years, {\it i.e.} since the $\nu \bar \nu$
pair production process has been implemented into KORALZ for the first
time \cite{colas}, the program was modified several times, and in
general its applicability for $s$-channel processes was assured also at
LEP2 energies, see \cite{ybook1}.

For the case $\nu_e \bar \nu_e \gamma $, however, the $WW\gamma$
interaction (and the question of gauge invariance) can no longer be
ignored, and the approximate treatment of $t$-channel $W$ exchange
requires refinements to be precise enough at LEP2 energies.

\begin{figure}[htb]
%\begin{center}
%\epsfig{file=geng2.eps,width=10cm,height=15cm}
 \unitlength 1mm
\begin{picture}(100,100)
%\lattice
  \put(20,-10){ \epsfxsize=10cm \epsfysize=15cm \epsfbox{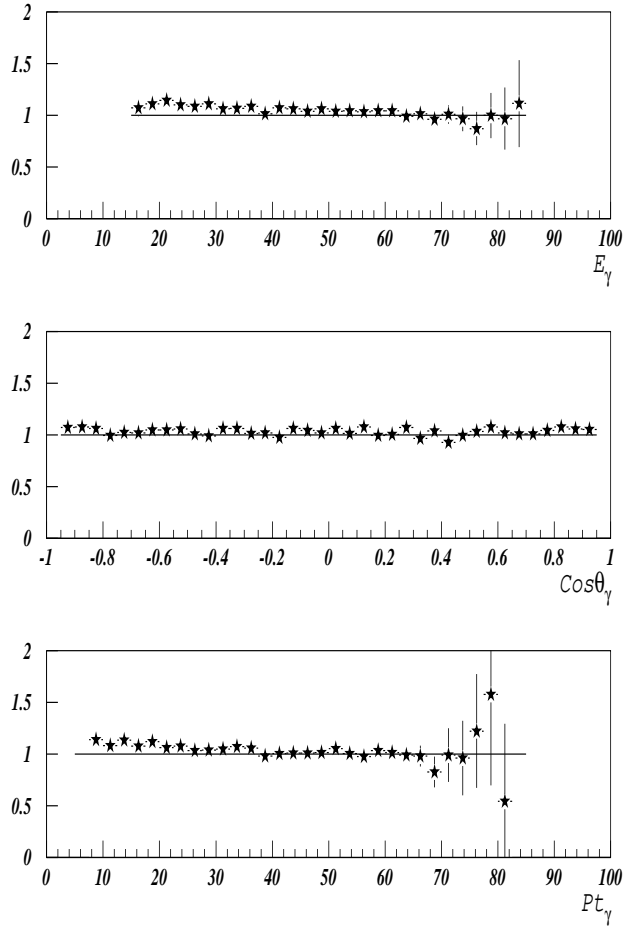}}
\end{picture}
\caption{ 
The ratio of distributions generated by the KORALZ 4.04
and the program used in Ref.~\protect\cite{debchou}. 
Both programs are run with the same input parameters and kinematical 
cuts. Anomalous $WW\gamma$ couplings are included.
\label{fig:technical}}
%\end{center}
\end{figure}

We now turn to the discussion of how these problems are solved in the
present version 4.04, which is the modification of the version 4.02 of
the KORALZ program.
The modification consists of introducing the
new weight defined as a ratio of matrix element squared for $e^+e^-
\to \nu \bar \nu \gamma$ with and without the $WW\gamma$
vertex:\footnote{In some regions of phase space, due to destructive
  interference of contributions from the $s$- and $t$-channel boson
  exchanges, the Standard Model predictions can be very small. There
  the weight defined in eq.~(\ref{weight}) can in principle tend to
  infinity, making the simulation numerically non-convergent.  This
  technical obstacle, not relevant to our discussion, is overcome by
  replacing the SM matrix element by a different one and obtaining the
  SM predictions also in the form of a weighted event sample.}
\begin{equation}
W_T=\frac{| {\cal M}_{0}+ {\cal M}_{WW\gamma}|^2}{|{\cal M}_{0}|^2}  .
\label{weight}
\end{equation}
The term ${\cal M}_{WW\gamma}$ contains contributions from diagram (5)
with the $WW\gamma$ vertex, including anomalous couplings. The matrix
element is taken from the numerical program developed in
Ref.~\cite{debchou}. Note that the term in the denominator $|{\cal
M}_{0}|^2$ is gauge dependent. If in the present version KORALZ 4.04
the events were generated according to the matrix element $|{\cal
M}_{0}|^2$, the gauge dependence would cancel out in the product of
eqs.~(\ref{Factor}) and (\ref{weight}). However, this is not the case,
since they are generated according to the approximate $|{\cal
M}_{0}^{approx} |^2$ as explained above. Nevertheless, as we will
discuss in the next chapter, see Fig.~\ref{fig:technical}, the
numerical error is small and the final distributions are affected by
this approximation in a small and {\it controlled} manner only.
The implementation of the above weight $W_T$ proceeds as follows.  For
every event, the complete kinematic configuration is taken.  In cases
when there is just one photon, the formula (\ref{weight}) can be used
in a direct way.  However, for events with multiple photons generated,
there are many possibilities for a choice of a photon radiated off the
$W$ boson.  Since the QED bremsstrahlung of high-$p_T$ photons from
electrons is strongly suppressed, and the $WW\gamma$ contribution is
important for energetic photons, in our algorithm we
calculate\footnote{ At this point we assume that the $WW\gamma$
  couplings do not affect the matrix elements in soft or collinear
  photon limits, or virtual corrections.  In short, it
  means that the whole structure of QED remains intact and that the
  sole effect of new couplings is a modification of the amplitude for
  configurations with photons of high energy and $p_T$.} the
weight factor $W_T$ in eq.~(\ref{weight}) only for the photon of 
highest $p_T$.  All
remaining photons are then
incorporated into effective beams according to the following recipe:\\
a) Except for the photon of highest $p_T$, all remaining photon's
four-momenta are subtracted from the four-momentum of the beam of the
same hemisphere. In this way we obtain
effective electron beams, which are not necessarily on-mass-shell.\\
b) We boost all four-momenta of effective beams, neutrinos as well as
the highest-$p_T$ photon to the rest frame of effective beams.
\\
c) We reset the four-momenta of the effective electron beams to be
on-mass-shell in such a way that their back-to-back direction and
energy-momentum conservation is preserved.

At this point we have constructed reduced on-mass-shell kinematics for 
 $e^+e^- \to \nu \bar \nu \gamma$ configuration for which the weight 
$W_T$  can be calculated according to eq.~(\ref{weight}).
For practical reasons explained at the end of the next chapter,
we do not include the $W_T$ weight into KORALZ internal rejection loop.
Instead, generated sample follows the same distributions as in previous
versions of the program. The $WW\gamma$ contribution manifests itself
only through the weight $W_T$ which is provided for every event.

\vskip 0.4 cm
\centerline{\bf \large 4. Numerical tests}
\vskip 0.3 cm

The reduction procedure presented above is not free of systematic
uncertainties for $WW\gamma$ contributions to multiphoton production
processes, except at the single photon level.  These uncertainties,
though, are beyond the leading-log QED level as well as beyond 
soft-photon corrections.  Only genuine next-to-leading-log corrections to
contributions from $WW\gamma$ interactions are out of
control.  On the other hand, one benefits from full coverage of the
bremsstrahlung photons phase space provided by the host Monte Carlo as
well as its predictions for Standard Model without any deterioration
in quality.

As a first step in testing the implementation of the $WW\gamma$ vertex
(with or without anomalous couplings) and diagrams including
$t$-channel $W$-exchange into KORALZ, we performed studies at a
single-photon level. In this case results from KORALZ should be the
same as from direct Monte Carlo integration (using weighted event
sample), as {\it e.g.} in the original approach of \cite{debchou}. Any
discrepancy would indicate either technical problems in any of the two
programs, or breakdown of our approach.  The results are presented in
Fig.~\ref{fig:technical}, where the ratios of differential cross
sections obtained from two programs are shown. The results are
consistent over the entire phase space\footnote{For large 
$E_\gamma$ or $p_T^{\gamma}$ the cross section is very small.
Large statistical fluctuations in  the ratios 
of the  Monte Carlo results are thus expected.}. 
One can see that  distributions 
differ by at most a few percent. This also shows that the
numerical error due to breaking the gauge invariance by our procedure
is small.

In the second step we verify that the reduction procedure to effective
beams as explained in the previous section indeed works. For that
purpose we compare at a given cms (center-of-mass system) 
energy the results from KORALZ with
anomalous contributions and higher-order corrections included as
described above, with another approach (called EMU), 
where convolution of the $WW\gamma$ 
amplitude with higher-order corrections is performed as explained below. In
the EMU approach, we first generate the spectrum of $M_{\it
 eff}$ from the multiphoton run of KORALZ at  cms energy
without any $WW\gamma$ interaction.  The $M_{ {\it eff}}$ is the
invariant mass of the $\nu \bar \nu \gamma_{p_T^{max}}$ system. For
practical reasons we divide the $M_{{\it eff}}$ spectrum into discrete
bins.  In the second step, for each bin $i$ containing $N_i$ events we
generate an equal number of events, this time, however, at ``an effective
cms energy'' equal to the centre of the bin $M_{{\it eff}}$ with the
help of a single bremsstrahlung mode of KORALZ operation and
contributions from the $WW\gamma$ interaction included (which we have
shown before to work properly).  In this case additional photons are
not present, but we emulate their presence by a boost of all 
final-state particles from the ``effective cms energy'' $M_{{\it eff}}$ to
the overall cms frame as if there was (just one) additional photon
collinear to one of the beams.

\begin{figure}[htb]
 \unitlength 1mm
\begin{picture}(100,100)
%\lattice
  \put(20,-10){ \epsfxsize=10cm \epsfysize=15cm
\epsfbox{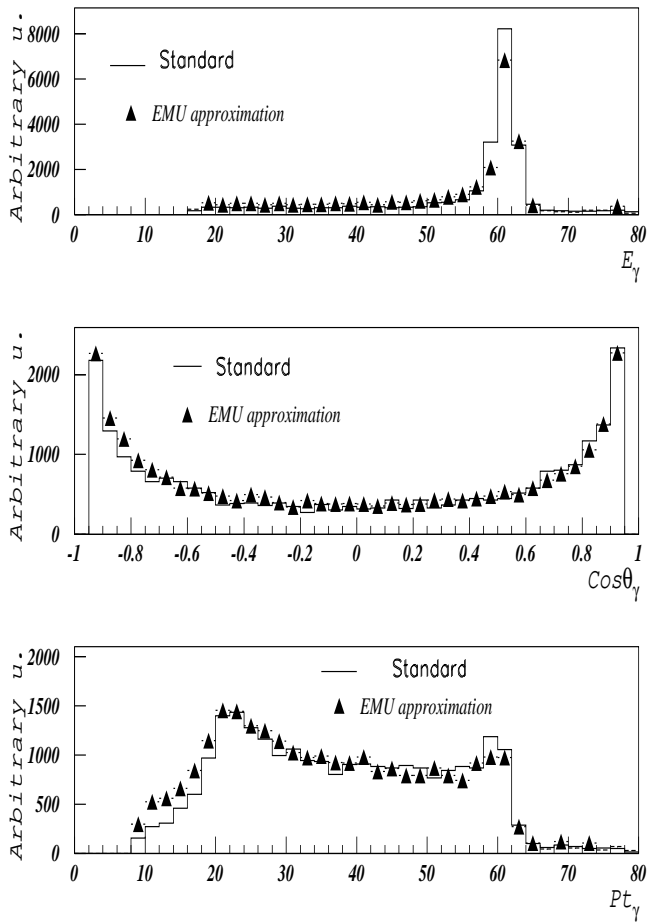}}
\end{picture}
%\begin{center}
%\epsfig{file=comp3.eps,width=10cm,height=15cm}
\caption{Comparison of differential distributions
obtained from the new  version of KORALZ 4.04
including anomalous interactions (Standard),  and the EMU
algorithm (triangles), as  
explained in the text.
\label{fig:consistency}}
%\end{center}
\end{figure}

As can be seen in Fig.~\ref{fig:consistency}, the two plots differ in
the physically interesting region\footnote{ Physically interesting
region consists in this case of photons of $p_T> 20 $~GeV and of
energies different from radiative return to the $Z$.} by only up to a
few per cent. This indicates that the proposed algorithm is
constructed properly and also that the bulk of the higher-order
corrections is of leading-log type.

One should bear in mind that such tests are indicative only.  To
answer the question of how good such an algorithm is, would require
careful comparisons of the generated distributions with those from
exact matrix-element calculations including anomalous couplings of order 
at least  $\alpha^2$.  This is difficult and not necessary at the
moment, as a precision on higher-order corrections to anomalous
contributions of order 10\% is sufficient.  

In practical applications it is convenient to use the method of
correlated samples.  The $WW\gamma$ vertex is linear in \dek and 
$\lambda_{\gamma}$,
the two independent anomalous couplings.  Therefore the matrix element
squared, and consequently all differential distributions, are bilinear
in the variables \dek and $\lambda_{\gamma}$.  It is thus convenient
to calculate, for every generated point in phase-space, the value of the
matrix element at six fixed and conveniently chosen combinations of
\dek and \lam values\footnote{ In KORALZ numerical values of 
($\Delta\kappa_{\gamma}$,  $\lambda_{\gamma}$) actually implemented are:
(0, 0), (-10, 0), (10, 0), (0, -10), (0, 10), (10, 10). }. 
{}From this, the matrix element for any other value
of anomalous couplings can be easily recalculated.  This property
extends to differential distributions as well. It also remains valid
when higher-order leading-log QED corrections as explained above are
included.

Note that KORALZ generates unweighted events for the Standard Model with
additional weights corresponding to different modifications of the
matrix element due to $WW\gamma$ couplings. This not only saves
computing time, as it is sufficient to perform detector simulation
only once, but also the standard and non-standard model samples
obtained this way are statistically correlated,  as was explained in
\cite{AFB}.  This significantly improves the convergence of the
predictions since the statistical error affects only the difference of
the two distributions.  

After these modifications, KORALZ version 4.04 (or higher) can be
used\footnote{The version 4.04 can be obtained upon request from
Zbigniew.Was@cern.ch. It includes also options described in
\cite{LEP1}.} for our purpose with any combination of flags for the
initial state. Every event generated as weight-one event,
corresponding to the Standard Model with $WW\gamma$ vertex
excluded\footnote{Exactly as in the versions of the program prior to
the modifications presented here.}, is accompanied now by the weight
record, allowing for a calculation of weights for any numerical value
of $WW\gamma$ coupling, standard or anomalous. This can be especially
useful in performing final fits to the data in the presence of
experimental cuts. It can also be used for Standard Model background
calculations for other processes with a single photon plus missing
energy in the final state.

Finally, let us note that the question of systematic errors
of the genuine $s$-channel process  for non-electron neutrino pair 
production was not addressed. It calls for a separate discussion 
including tests/comparisons with, for instance, the matrix element 
for the production of a neutrino pair with up to three real protons.

\vskip 0.4 cm
\centerline{\bf \large 5. Results and Conclusions}
\vskip 0.3 cm

\begin{figure}[htb]
 \unitlength 1mm
\begin{picture}(100,100)
%\lattice
  \put(20,-10){ \epsfxsize=10cm \epsfysize=12cm
\epsfbox{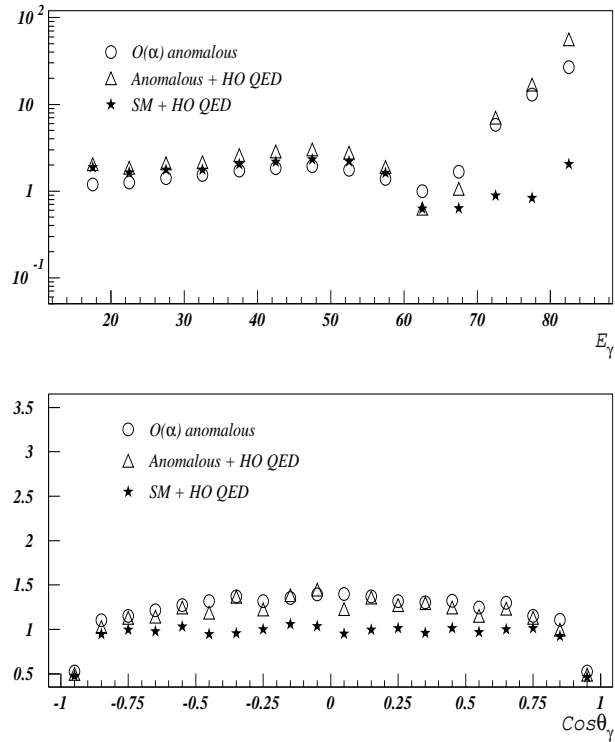}}
\end{picture}
%\begin{center}
%\epsfig{file=gener.eps,width=10cm,height=14cm}
\caption{The ratios of $E_\gamma$ and $\cos \theta_\gamma$ distributions
  obtained from: (i) order-$\alpha$ distributions including anomalous
  couplings -- open circles, (ii) multiple photon radiation, no
  anomalous couplings -- stars, (iii) multiple photon radiation and
  anomalous couplings included -- triangles. In all cases
  distributions are divided by the appropriate ones obtained from the
  single-photon generation in the Standard Model (no anomalous
  couplings).  All input parameters as explained in the text.  
\label{fig:ratios}}
%\end{center}
\end{figure}

Let us now turn to the presentation of the numerical results for
anomalous couplings. In Fig.~\ref{fig:ratios} we compare the photon
spectra normalized to the SM lowest-order calculations for three
cases: (i)~anomalous couplings are included at the lowest order
(circles), (ii)~higher-order QED corrections but no anomalous
couplings are included (stars), and (iii)~both anomalous couplings and
higher-order corrections are included (triangles). The difference of
physical content between this plot and the one of
Fig.~\ref{fig:classical} consists primarily in presenting the results
for the case where higher-order radiative corrections and anomalous
contributions are included simultaneously. Although the QED
higher-order bremsstrahlung corrections are sizeable, and varying from
$-$30\% to 100\%, they do not wash out the effect of the anomalous
contribution. What is more interesting, in the region of high-energy
photons the presence of higher-order effects tends to slightly enhance
the sensitivity to anomalous couplings. On the other hand, for photon
energies below that of radiative return, the sensitivity deteriorates
significantly. We also observe that the combined effect of anomalous
couplings and radiative corrections is not just a simple sum of the
two effects.

We have thus shown that the interpretation of experimental data for
anomalous $WW\gamma$ couplings in the $e^+e^- \to \nu \bar \nu \gamma$
channel requires theoretical predictions in the form of a Monte Carlo
simulation program, which includes  anomalous contributions
as well as multiple bremsstrahlung higher-order QED corrections. The
current version 4.04 of the KORALZ program, with modifications presented in
this paper, can now be used to achieve this goal. It can also be used 
for Standard Model background calculations for other interesting
processes, such as gravitino or neutralino/sneutrino production 
processes at LEP2.

\vspace{4mm}
\noindent {\bf \large Acknowledgements}\\

This work has been supported in part by the Polish-French
Collaboration within IN2P3 (AJ), by the Polish State
Committee for Scientific Research grant 2 P03B 030 14 (JK), 
and by  the II Maria Sk\l odowska-Curie Fund
PAA/DOE-97-316 (ZW).  ZW acknowledges the hospitality and support of the ALEPH
Collaboration group in LAL Orsay. JK thanks the CERN Theory Division
for hospitality during the final stage of this work.

%%%%%%%%%%%%%%%%%%%%%%%%%%%%%%%%%%%%%%%%%%%%%%%%%%%%%%%%%%%%%%%%%%%%%%%%%%%%
%%%%%%%%%%%%%%%%%%%%%%%%%%%%%%%%%%%%%%%%%%%%%%%%%%%%%%%%%%%%%%%%%%%%%%%%%%%%
%%%%%%%%%%%%%%%%%%%%%%%%%%%%%%%%%%%%%%%%%%%%%%%%%%%%%%%%%%%%%%%%%%%%%%%%%%%%

\end{document}